\documentclass{article}
\usepackage{epsfig}
\usepackage{amssymb}

\date{\today}

\usepackage{amsmath}

\newcommand{\al}{\alpha}

\newcommand{\ee}{\end{equation}}
\newcommand{\eea}{\end{eqnarray}}
\newcommand{\be}{\begin{equation}}
\newcommand{\bea}{\begin{eqnarray}}

\newcommand{\re}[1]{(\ref{#1})}

\begin{document}

\begin{center}
{\Large\bf New  AdS non Abelian black holes with superconducting horizons }
\vspace{0.5cm}
\\
{\bf Ruben Manvelyan$^{1,2}$},
{\bf Eugen Radu$^{3}$ }
 and  {\bf D. H. Tchrakian $^{4}$ }

\vspace*{0.2cm}
{\it $^1$  Yerevan Physics institute, Alikhanian Br. St. 2, 0036 Yerevan, Armenia}
\\
$^{2}${\it  Department of Mathematical Physics, National University of
Ireland, Maynooth, Ireland}
\\
{\it $^3$ Institut f\"ur Physik, Universit\"at Oldenburg, Postfach 2503
D-26111 Oldenburg, Germany}
\\
{\it $^{4}$School of Theoretical Physics -- DIAS, 10 Burlington
Road, Dublin 4, Ireland}
\vspace{0.5cm}
\end{center}
\begin{abstract}
We present arguments for the
existence of higher dimensional asymptotically AdS non Abelian black holes with a Ricci flat event horizon
and analyze their basic properties.  Unlike higher dimensional black holes with a curved horizon, of the usual
Einstein-Yang-Mills system, these solutions have finite mass-energy.
Below some non-zero critical temperature, they are thermodynamically preferred over the Abelian configurations.
\end{abstract}
%%%%%%%%%%%%%%%%%%%%%%%%%%%%%%%%%%%%%%%%%%%%%%%%%%%%%%%%%%%%%%%%%%%%%%%

%%%%%%%%%%%%%%%%%%%%%%%%%%%%%%%%%%%%%%%%%%%%%%%%%%%%%%%%%%%%%%%%%%%%%%%%%%%%%%%%%%%%%%%%%
%%%%%%%%%%%%%%%%%%%%%%%%%%%%%%%%%%%%%%%%%%%%%%%%%%%%%%%%%%%%%%%%%%%%%%%%%%%%%%%%%%%%%%%%%
%%%%%%                      INTRODUCTION
%%%%%%%%%%%%%%%%%%%%%%%%%%%%%%%%%%%%%%%%%%%%%%%%%%%%%%%%%%%%%%%%%%%%%%%%%%%%%%%%%%%%%%%%%
%%%%%%%%%%%%%%%%%%%%%%%%%%%%%%%%%%%%%%%%%%%%%%%%%%%%%%%%%%%%%%%%%%%%%%%%%%%%%%%%%%%%%%%%%
\section{Introduction}

In recent years it became clear that various well-known, and rather intuitive, features of
self-gravitating solutions with Maxwell fields in $d=3+1$ spacetime dimensions
are not shared by their counterparts with non Abelian gauge
fields. For example, in contrast to the Abelian situation,
self-gravitating Yang-Mills (YM) fields can form particle-like configurations
\cite{Bartnik:1988am}. The Einstein-Yang-Mills (EYM)
equations also admit black hole solutions that are not uniquely
characterised by their mass, angular momentum  and YM charges
\cite{Volkov:sv}. Therefore the uniqueness theorem for electrovacuum
black hole spacetimes  ceases to exist for EYM systems.
As a result, the literature on  gravitating solutions with non Abelian
fields has steadily  grown up in the last two decades,
including solutions with a cosmological constant $\Lambda$
(see e.g. \cite{Mann:2006jc,Winstanley:2008ac} and references therein).
The asymptotically anti--de Sitter (AAdS) solutions are of particular interest,
since gauged supergravity theories playing
an important role in AdS/CFT, generically contain non Abelian matter fields in
the bulk, although to date mainly Abelian truncations are considered in the literature.
Notably, non Abelian AAdS solutions exhibit new features which are absent for $\Lambda=0$.
For example, stable solutions with a global magnetic charge are known to exist
even in the absence of a Higgs field \cite{Winstanley:1998sn}, \cite{Bjoraker:2000qd}.

In the context of the AdS/CFT correspondence~\cite{AdS/CFT}, Klebanov and Witten have proposed a
mechanism of spontaneously breaking gauge symmetry~\cite{Klebanov}. This mechanism has recently been exploited by
Gubser {\it et. al.}~\cite{Gubser:2008zu}--\cite{Abelian}
to explain important phenomena in condensed matter physics, in particular superconductivity and critical
phenomena. This mechanism results, with no recourse to supersymmetry, in a symmetry breaking boundary theory of a
bulk gravitational theory with negative cosmological constant, the temperature of the
black hole being nonzero.

From our point of view, the most interesting development in this domain is the recent discovery in \cite{Gubser:2008zu}
that some AAdS non Abelian black hole solutions with a Ricci flat event horizon may posses
superconducting horizons which are thermodynamically preferred below some non-zero critical temperature.
Such solutions exhibit hair of
the 'electric' part of the gauge field on the AdS boundary, manifesting the gauge symmetry breaking mechanism; at the
same time the condensate of the 'magnetic' part floats above the horizon of the black hole.
This mechanism was further exploited in subsequent works~\cite{Gubser:2008wv,Roberts:2008ns}.

The only case discussed so far pertain to four dimensional AAdS spacetimes  with Ricci flat horizon, and
relatively little is known about  such  higher dimensional
solutions with non Abelian matter fields. Naturally, it is always of interest to see how the dimensionality of the
spacetime affects the physical consequences of a given theory. In particular, it would be interesting to
see how general is the mechanism discovered in \cite{Gubser:2008zu}. Besides, it is known from the work of
\cite{Okuyama:2002mh}, \cite{Volkov:2001tb}, \cite{Brihaye:2005tx}, \cite{Radu:2005mj}, that static
spherically symmetric solutions of the usual gravitating YM system
in spacetime dimensions $d>4$  do not have finite energy as a result of their scaling propertes.
Finite energy solutions exist only when the
usual YM system is augmented with higher derivative corrections
in the non Abelian action~\cite{Radu:2005mj}, \cite{Brihaye:2002hr}. Therefore
the examination of higher dimensional gravitating
non Abelian solutions with a different topology (in this case with Ricci flat)
of the event horizon is a pertinent task.

Our objective in the present work is to extend this type of symmetry breaking to non Abelian EYM solutions used by
Gubser in \cite{Gubser:2008zu}, to arbitray $d=D+1$ dimensional AAdS spacetimes. For this we exploit, qualitatively,
previous results on $d$ dimensional finite energy AAdS solutions for EYM systems given in \cite{Radu:2005mj}. In fact,
the actual model considered here differs from those of \cite{Radu:2005mj} in that the latter is described by the purely
'magnetic' components of the YM field\footnote{Inclusion of the 'electric' components of the YM field can readily be made,
as {\it e.g.} in the case of Euclidean signature in \cite{Radu:2007az}.} whereas here our model will include also the
'electric' components which play an essential role as is the case of \cite{Gubser:2008zu}. The salient feature of the YM
models in \cite{Radu:2005mj} is the presence of higher order terms in the YM curvature, whose role is to supply
the necessary (Derick) scaling
properties of the Lagrangian to enable the existence of finite energy solutions. There is however one major
departure between the models in \cite{Radu:2005mj} and those exploited here. While the metric Ansatz employed in
\cite{Radu:2005mj} describes a spacetime with an $S^{d-2}\times R$ boundary at infinity, the one here describes in contrast a flat
Minkowkian boundary. One consequence of this is that the appropriate gauge group here is $SO(D-1)$,
{\it i.e.} $SO(d-2)$, differing from the choice of $SO(D)$, {\it i.e.} $SO(d-1)$, in \cite{Radu:2005mj}.
This feature is a reminder of the fact that the electric component $A_0$ of the YM connection takes the role of a
Higgs field. The other consequence of a flat
Minkowkian boundary is that inclusion of higher order YM curvature terms are no
longer necessary for the solutions to describe finite energy configurations,  as was the case when the black hole horizon
had a nonzero Ricci tensor. This is a result of the much wider range of
scaling properties satisfied when the metric has a Ricci flat event horizon instead of the more restrictive scaling
properties of the system when the horizon is spherical. Inclusion of higher order YM curvature terms, while not
necessary for achieving finite energy, is still possible here, resulting only in quantitative effects.   We have
eschewed this option here since it is not qualitatively important.
In addition, although we have restricted our attention here to spacetime dimensions $5\le d\le 8$ for simplicity,
it is obvious that this limitation is unimportant.

The metric Ansatz we use is a direct extension of that in \cite{Winstanley:1998sn,Gubser:2008zu,Radu:2002hf}, to
dimensions with a larger number of spacelike coordinates. To implement our procedure it is necessary to devise an Ansatz
for the YM connection, generalising that used in previous work on YM fields in AdS spacetime. Here, we have found
two distinct Ans\"atze which we have verified to be consistent. These generalise the distinct YM connection Ans\"atze
of \cite{Gubser:2008zu} and of \cite{Gubser:2008wv}, respectively.

%%%%%%%%%%%%%%%%%%%%%%%%%%%%%%%%%%%%%%%%%%%%%%%%%%%%%%%%%%%%%%%%%%%%%%%%%%%%%%%%%%%%%%%%%
\section{General formalism}
%%%%%%%%%%%%%%%%%%%%%%%%%%%%%%%%%%%%%%%%%%%%%%%%%%%%%%%%%%%%%%%%%%%%%%%%%%%%%%%%%%%%%%%%%
%%%%%%%%%%%%%%%%%%%%%%%%%%%%%%%%%%%%%%%%%%%%%%%%%%%%%%%%%%%%%%%%%%%%%%%%%%%%%%%%%%%%%%%%%
\subsection{The field equations and the abelian solution}
%%%%%%%%%%%%%%%%%%%%%%%%%%%%%%%%%%%%%%%%%%%%%%%%%%%%%%%%%%%%%%%%%%%%%%%%%%%%%%%%%%%%%%%%%
Instead
of specializing to a particular supergravity model, we shall consider the
pure EYM theory with negative cosmological
constant in $d\geq 4$ spacetime dimensions
\begin{eqnarray}
\label{action}
S=\int d^dx\sqrt{-g}~\left(\frac{1}{16 \pi G}(R-2 \Lambda)-\frac{1}{4} \,F_{\mu \nu}^aF^{a\mu \nu}  \right),
\end{eqnarray}
where the cosmological constant is $\Lambda=-(d-2)(d-1)/2\ell^2$.
Although it seems that the model (\ref{action}) is non-supersymmetric
in itself (at least\footnote{The case $d=4$, with $\Lambda/(16 \pi G)=-3 g^2$
corresponds to a consistent truncation of ${\cal N}=4$ gauged supergravity and may be uplifted
to $d=11$ supergravity \cite{Pope:1985bu},
\cite{Cvetic:1999au},\cite{Mann:2006jc}.} for $d>4$), it usually enters the gauged supergravities as the basic
building block. Therefore  one can expect the basic features of its solutions to be
generic.

Variation of the action (\ref{action})
 with respect to the metric $g^{\mu\nu}$ and the gauge field $A_\mu$ leads to the EYM equations
\begin{equation}
\label{einstein-eqs}
R_{\mu\nu}-\frac{1}{2}g_{\mu\nu}R +\Lambda g_{\mu\nu}  = 8 \pi G T_{\mu\nu},~~D_{\mu}F^{\mu\nu}=0,
\end{equation}
where $T_{\mu\nu}$  is the YM stress-energy tensor
%\begin{eqnarray}
%\label{tik}
$T_{\mu\nu} =
  F_{\mu\alpha}^a F_{\nu\beta}^a g^{\alpha\beta}
   -\frac{1}{4} g_{\mu\nu} F_{\alpha\beta}^a F^{a\alpha\beta} ,$ 
and $D_{\mu}=\partial_{\mu} + ig \left[A_\mu , \cdot \right]$
(with $g$ the gauge coupling constant).

We shall consider black hole
solutions of the above equation with locally flat horizons, which asymptotically
approach a locally AdS spacetime  with a boundary at infinity $R^{d-1}$.
The simplest such configuration with a nonzero gauge field is represented by the Reissner-Nordstr\"om-AdS (RNAdS) black hole with
\begin{eqnarray}
\label{RNAdS}
&ds^2_d=\frac{dr^2}{\frac{r^2}{\ell^2} -\frac{2M_0}{r^{d-3}} +\frac{8 \pi G(d-3)}{g^2(d-2)}\frac{q^2}{r^{2(d-3)}}}
 +  r^2d\gamma^2 -
 (\frac{r^2}{\ell^2} -\frac{2M_0}{r^{d-3}} +\frac{8 \pi G(d-3)}{g^2(d-2)}\frac{q^2}{r^{2(d-3)}})  dt^2,
~~
 A=(c-\frac{1}{g} \frac{q}{r^{d-3}}){\cal T }dt,
\end{eqnarray}
where $M_0,c,q$ are constants, ${\cal T }$ is an element of the gauge  group while
 $d\gamma^2$ is the line element of the $d-2$ euclidean space.
 The parameters $M_0$ and $q$ are proportional to the mass and electric charge of the solution.
 The black hole horison is located at $r=r_h$, with $ {r_h^2}/{\ell^2} -{2M_0}/{r_h^{d-3}}
 +({8 \pi G(d-3)}/{g^2(d-2)} ){q^2}/{r_h^{2(d-3)}}=0$.
When taking $q=0$, the Schwarzschild-AdS (SAdS) black hole with a planar horizon is recovered.

%%%%%%%%%%%%%%%%%%%%%%%%%%%%%%%%%%%%%%%%%%%%%%%%%%%%%%%%%%%%%%%%%%%%%%%%%%%%%%%%%%%%%%%%%
\subsection{The Ansatz}
%%%%%%%%%%%%%%%%%%%%%%%%%%%%%%%%%%%%%%%%%%%%%%%%%%%%%%%%%%%%%%%%%%%%%%%%%%%%%%%%%%%%%%%%%
We are interested in non Abelian configurations whose magnetic gauge potential
vanishes asymptotically, such that the abelian configuration (\ref{RNAdS}) is approached in that limit.
Moreover, we shall suppose that our configurations present a dependence only on a suitable radial coordinate $r$
which is orthogonal to the boundary of the spacetime.

The choice of the  gauge group compatible with these assumptions (and the corresponding YM Ansatz) is quite flexible.
For the $d=4$ case, two different\footnote{This contrasts with the case of solutions with
a $R\times S^{d-2}$ boundary  at infinity, where the choice of  Ansatz is unique, see $e.g.$ \cite{Bjoraker:2000qd}.}
non Abelian Ans\"atze have been proposed
 in the literature, both of them for a gauge group SU(2).
The first Ansatz used in  \cite{Gubser:2008zu} corresponds to a "circular polarisation" of the
magnetic YM connection and leads to an isotropic energy momentum tensor for the components on a surface of constant
$(r,t)$, with $t$ the time coordinate.
This is not the case for the second YM Ansatz proposed in \cite{Gubser:2008wv}, where
a particular direction in the $R^2$ subspace is choosen, leading to a more complicated metric Ansatz.

A straightforward generalisation of the isotropic Ansatz in  \cite{Gubser:2008zu}
is found for a gauge group $SO(D)$ (with $D<d$), the Ansatz for the YM connection being stated by
\begin{eqnarray}
A(r)&=&u(r)n^{i}_{\alpha}\Gamma_{ij}m^{j}_{\alpha} dt +w(r)dx^{i}\Gamma_{iD},
\label{AnsatzI}
\end{eqnarray}
in which $\Gamma_{AB}=(\Gamma_{ij},\Gamma_{iD})$ are the gamma matrices in $D$ dimensions and the indices $\al,\beta$
run over the range $\alpha, \beta =1,2,..(D-1)/{2}$. In the above relation,   $x^i$
are the coordinates parametrizing  a surface of constant $r,t$.
%Clearly, this restricts us to even values of $d-n$ only.
Also, the sets $(n^{i}_{\alpha}, m^{j}_{\alpha})$ appearing in \re{AnsatzI}  form a complete and orthonormal basis of
constant valued vectors of unit length,
\begin{eqnarray}
\nonumber
\sum^{D-1}_{i=1} n^{i}_{\alpha}m^{i}_{\beta}~= 0,
~~
\sum^{D-1}_{i=1} n^{i}_{\alpha}n^{i}_{\beta} = \delta_{\alpha\beta}\quad,\quad
\sum^{D-1}_{i=1} m^{i}_{\alpha}m^{i}_{\beta} = \delta_{\alpha\beta},
\label{cc2}
~~
\sum^{\frac{D-1}{2}}_{\alpha=1} n^{i}_{\alpha}n^{j}_{\alpha}
+ \sum^{\frac{D-1}{2}}_{\alpha =1} m^{i}_{\alpha}m^{j}_{\beta} = \delta^{ij}.
\end{eqnarray}
The last completeness condition is very important and means that $D-1$ is even.
Therefore, for $d=D+1$, the YM Ansatz (\ref{AnsatzI}) is
defined\footnote{The construction of a consistent isotropic Ansatz for $d=2k+1$ necessitates
the enlargement of the gauge group, resulting in a YM connection with more than one magnetic potential.
For example for the most interesting $d=5$ case,
the minimal gauge group is SO(5), while the Ansatz would contain six non Abelian potentials.} only for an even number of
space dimensions $d$. Of course, it can also be used for an arbitrary spacetime dimension by adding $n$ codimensions $y^k$,
with the non Abelian potential identically zero in  the subspace labeled by the extra-coordinates.
Obviously, this Ansatz leads to an isotropic energy momentum tensor in both $x^i$ and $y^k$-directions,
with $T_{x^i}^{x^i}\neq T_{y^k}^{y^k}$.

However, it is possible to define a different non Abelian Ansatz (Ansatz II in what follows),
valid for all even and odd dimensions $d$. In this case we
employ only one (constant valued) set of orthonormal unit vectors $n^{i}$, in terms of which
\begin{eqnarray}
&&A(r)=u(r)n^{i}\Gamma_{i,d-1}dt +w(r)n^{j}\Gamma_{ji}dx^{i}~.\label{AnsatzII}
\end{eqnarray}
While the components of the connection \re{AnsatzII}  take their values fully in the algebra of $SO(d-1)$, only
$d-2$ components of the YM connection are effective since $n^{i}A_{i}=0$. Thus we have one additional radius,
but the fields depend only on the radial coordinate $r$, so one of the magnetic components is zero.
This results also on an anisotropic energy mementum tensor of the YM field.

For an arbitrary spacetime dimension, a metric form compatible with the above two YM ans\"atze is given by
\begin{eqnarray}
\label{metric}
ds^2_d=A(r) dr^2 +F_1(r) \sum_{i=1}^{d-n-2}dx^idx^i+F_2(r) \sum_{k=1}^{n}dy^idy^i -B(r)  dt^2,
\end{eqnarray}
where we have found convenient to take
\begin{eqnarray}
\label{N}
A(r)=\frac{1}{N(r)},~~~B(r)=N(r)\sigma^2(r),~~~F_1(r)=r^2f^2(r),~~~F_2(r)= r^2(f(r))^{2(n+2-d)/n}
\end{eqnarray}
with
\begin{eqnarray}
\label{N1}
N(r)=-\frac{2m(r)}{r^{d-3}}+\frac{r^2}{\ell^2},
\end{eqnarray}
 the function $m(r)$ being related to the local mass-energy density up to some
$d-$dependent factor.
For the case of the first Ansatz (\ref{AnsatzI}), the YM field is defined on a subspace labeled by the
$(r,t;x^i)$-coordinates, where $i= 1,\dots,d-n-2$ (with $n\geq 0$). The second YM ansatz corresponds to an arbitrary
$d$, with $n=1$ (Here we assume without any loss of generality $n^i=\delta_{d-2}^i$ and write $x^{d-2}=y^1$.)

%%%%%%%%%%%%%%%%%%%%%%%%%%%%%%%%%%%%%%%%%%%%%%%%%%%%%%%%%%%%%%%%%%%%%%%%%%%%%%%%%%%%%%%%%
\subsection{The equations of motion and asymptotic solutions}
%%%%%%%%%%%%%%%%%%%%%%%%%%%%%%%%%%%%%%%%%%%%%%%%%%%%%%%%%%%%%%%%%%%%%%%%%%%%%%%%%%%%%%%%%
Within these Ans\"atze, the EYM field equations reduce to a set of five ordinary differential
equations which can be expressed in a unified form as\footnote{Here and in what follows, the relations for the case
with no codimensions are found by formally setting $f\equiv 1$, followed by the limit $n\to 0$.} (where one takes
$c=1/2$ for the Ansatz I and $c=1/(d-3)$ for Ansatz II)
\begin{eqnarray}
\nonumber
&&w''=\left(
\frac{2f'}{f}-\frac{d-4}{r} -\frac{\sigma'}{\sigma} -\frac{N'}{N}
\right)w'
+\left(
\frac{u^2}{N\sigma^2}-\frac{(d-n-3)w^2}{r^2f^2}
\right)\frac{w}{N}=0,
\\
\nonumber
&&u''+\left(\frac{d-2}{r}-\frac{\sigma'}{\sigma} \right)u'-\frac{1}{c}\frac{ u w^2}{r^2Nf^2}=0,
\\
\label{eqs}
&&m'=\frac{d-n-2}{2n}r^{d-2}N\frac{f'^2}{f^2}
+\alpha^2 r^{d-4}
\left
(\frac{Nw'^2}{f^2}+c\frac{r^2u'^2}{ \sigma^2}+\frac{(d-n-3)w^4}{2r^2f^4}+\frac{u^2w^2}{N\sigma^2f^2}
\right),
\\
\nonumber
&&\sigma'=\frac{2\sigma}{f^2r }
\left(
\frac{d-n-2}{2n}r^2f'^2+
\alpha^2 ( w'^2+\frac{w^2u^2}{N^2\sigma^2})
\right),
\\
\nonumber
&&
f''=\frac{2n \alpha^2}{(d-n-2)r^2 f}
\left
(\frac{u^2 w^2}{N^2\sigma^2}-\frac{(d-n-3)w^4}{r^2f^2N}-w'^2
\right)
-
\left(\frac{d-2}{r}-\frac{f'}{f}+\frac{N'}{N}+\frac{\sigma'}{\sigma}
\right
)f',
%( \frac{r^{d-2}N\sigma}{f^2}f')'+\frac{f'^2}{f^3}
%+\frac{2n}{d-n-2}\alpha^2 r^{d-4}\sigma(\frac{Nw'^2}{f^2}
%+\frac{d-n-3}{r^2f^4w^4}+\frac{w^2\phi^2}{N\sigma^2f^2})=0,
\end{eqnarray}
where $\alpha^2=8\pi G/(g^2(d-2))$.
For $d=4\,,\ n=0$, the above equations reduce to those derived in \cite{Gubser:2008zu} albeit
for a different metric Ansatz\footnote{$d=4$ AAdS non Abelian black holes with a
Ricci flat horizon were discussed previously in \cite{Radu:2002hf}.}.
The Abelian Reissner-Nordstr\"om solution (\ref{RNAdS}) is found for
%\begin{eqnarray}
$m(r)=M_0- c(d-3)q^2 \alpha^2 /{ 2r^{d-3} },~~f(r)=\sigma(r)=1,~~w(r)=0,~~u(r)=u_0- {q}/{r^{d-3}}.$
%\end{eqnarray}

Unfortunately, no exact non Abelian solutions of this system are yet known.
However, one can analyse their properties by using a combination of analytical and numerical
methods, which are sufficient for most purposes.
The solutions have the following expansion\footnote{The case of Ansatz II for $d=4$ is special, as the
near horizon expansions of $w(r)$ and $f(r)$ are different in that case:
$w(r)=w_h+w_2(r-r_h)^2+O(r-r_h)^3\ $,
$\ f(r)=f_h+f_2(r-r_h)^2+O(r-r_h)^3$.
However, the horizon data is still determined by $\sigma_h,w_h,f_h,u_1$.}
at $r=r_h>0$ near the event horizon, located at $r_h$:
\begin{eqnarray}
\nonumber
&&m(r)=\frac{r_h^{d-1}}{2\ell^2}+m_1(r-r_h)+O(r-r_h)^2,
~
\sigma(r)=\sigma_h+\sigma_1(r-r_h)+O(r-r_h)^2,
\\
\label{b1}
&&f(r)=f_h+f_1(r-r_h)+O(r-r_h)^2,
\\
\nonumber
&&w(r)=w_h+w_1(r-r_h)+O(r-r_h)^2,
~~
u(r)=u_1(r-r_h)+u_2(r-r_h)^2+O(r-r_h)^3.
\end{eqnarray}
Here, following \cite{Gubser:2008zu}  we interpret $w_h$ as a 'magnetic' condensate.
All coefficients in the above relation can be
expressed in terms of the  real constants $\sigma_h,w_h,f_h,u_1$. One finds $e.g.$
\begin{eqnarray}
\nonumber
&&m_1= \frac{r_h^{d-6}\alpha^2}{2\sigma_h^2f_h^2}(2 cf_h^4u_1r_h^4+(d-n-3)\sigma_h^2w_h^4),
~~~~
f_1=-\frac{2\alpha^2(d-n-3)n r_h^{d-5}\ell^2w_h^2}{(d-n-2)f_h^2( (d-1)r_h^d-2m_1r_h^{2}\ell^2)},
\\
\label{b1-1}
&&\sigma_1= \frac{1}{f_h^2}\left(
\frac{2\sigma_h w_1^2 \alpha^2}{r_h}
+\frac{(d-n-2)}{n}f_1^2r_h\sigma_h
+\frac{2\alpha^2r_h^{2d-3}\ell^4u_1^2w_h^2}{((d-1)r_h^d-2m_1r_h^2\ell^2)^2\sigma_h}
\right),
\\
\nonumber
&&w_1=\frac{(d-n-3)\ell^2 w_h^3}{f_h^2 r_h^3(d-1-\frac{2m_1\ell^2}{r_h^{d-2}}) },
~~~~
u_2=\frac{u_1}{2}\left(-\frac{d-2}{2}+\frac{\sigma_1}{\sigma_h}
+\frac{ \ell^2w_h^2}{c f_h^2r_h^3(d-1-2m_1r_h^{2-d}\ell^2)}\right)\,.
\end{eqnarray}
Note also that the physical condition $N'(r_h)>0$
implies the following condition on the boundary data
%\begin{eqnarray}
$ {2c f_h^2u_1^2}/{\sigma_h^2}+ {(d-n-3)w_h^4}/{(r_h^4f_h^2)} < {(d-1)}/({\alpha^2\ell^2}).$
%\end{eqnarray}

We are interested in solutions of the EYM equations
approaching at infinity the Abelian RNAdS solution (\ref{RNAdS}).
This implies the following
 asymptotic expansion as $r \to \infty$
\begin{eqnarray}
\nonumber
&&m(r)=M_0- \frac{(d-3)\alpha^2}{ \ell^2}(J^2+ {c q^2 \ell^2} )\frac{1}{r^{d-3}}+O(1/r^{d-1}),
~
\sigma(r)=1-\frac{(d-3)^2}{(d-2)}\frac{J^2}{r^{2(d-2)}}+ O(1/r^{2(d+2)}),
\\
\label{b2}
&&f(r)=1- \frac{\bar f}{r^{d-1}}+ O(1/r^{2(d-2)}),
~~
w(r)=\frac{J}{r^{d-3}}+ O(1/r^{d-1}),
~~
u(r)=u_0-\frac{q}{r^{d-3}}+ O(1/r^{2(d-2)}),
\end{eqnarray}
with $M_0,J,q,\bar f$ real constants. The holographic interpretations of $u_0, q,$ and $J$ are   as follows:
$u_0$ is the chemical potential, $q$ is the electric charge, and $J$ is that component of the current $J_i$ on the
boundary connected with the spontaneously broken part of the bulk gauge symmetry \cite{Gubser:2008zu}. 
  In other words,   we have a  $D$-dimensional conformal field theory   described  on the boundary of
$AdS_{D+1}$ space equipped with $SO(D-1)$ conserved currents, which satisfy their own current algebra\footnote{The dual  CFT global
current $J_i$ is defined through $\mbox{Tr}A_iJ_i$, where
$A_i$ here is the asymptotic YM connection.}. 
 The  normalisable boundary value of the
'magnetic' field $w(r)= {J}/{r^{d-3}}+ O(1/r^{d-1})$ corresponds to the vacuum expectation value of the boundary
currents proportional to $J$ arising after symmetry breaking, and, the existence of the horizon 'magnetic' condensate
$w_{h}$.

The case $d=4,~n= 0$ considered in \cite{Gubser:2008zu} is special from the point of view of the
asymptotic expansion, since the finite energy requitements are
compatible with a nonvanishing value of the magnetic potential at infinity.
The condition $w(\infty)=0$ is imposed there by requiring $w$
to make a finite contribution to the norm of the non Abelian potential \cite{Gubser:2008zu}.
However, one can easily see from the field equation ({\ref{eqs}) that for $d>4$, $w(\infty)\neq 0$ results in a
divergent value of
the mass function $m(r)$, which gives further justification to the choice (\ref{b2}).

 %%%%%%%%%%%%%%%%%%%%%%%%%%%%%%%%%%%%
\begin{figure}[ht]
\hbox to\linewidth{\hss%
	\resizebox{7cm}{6cm}{\includegraphics{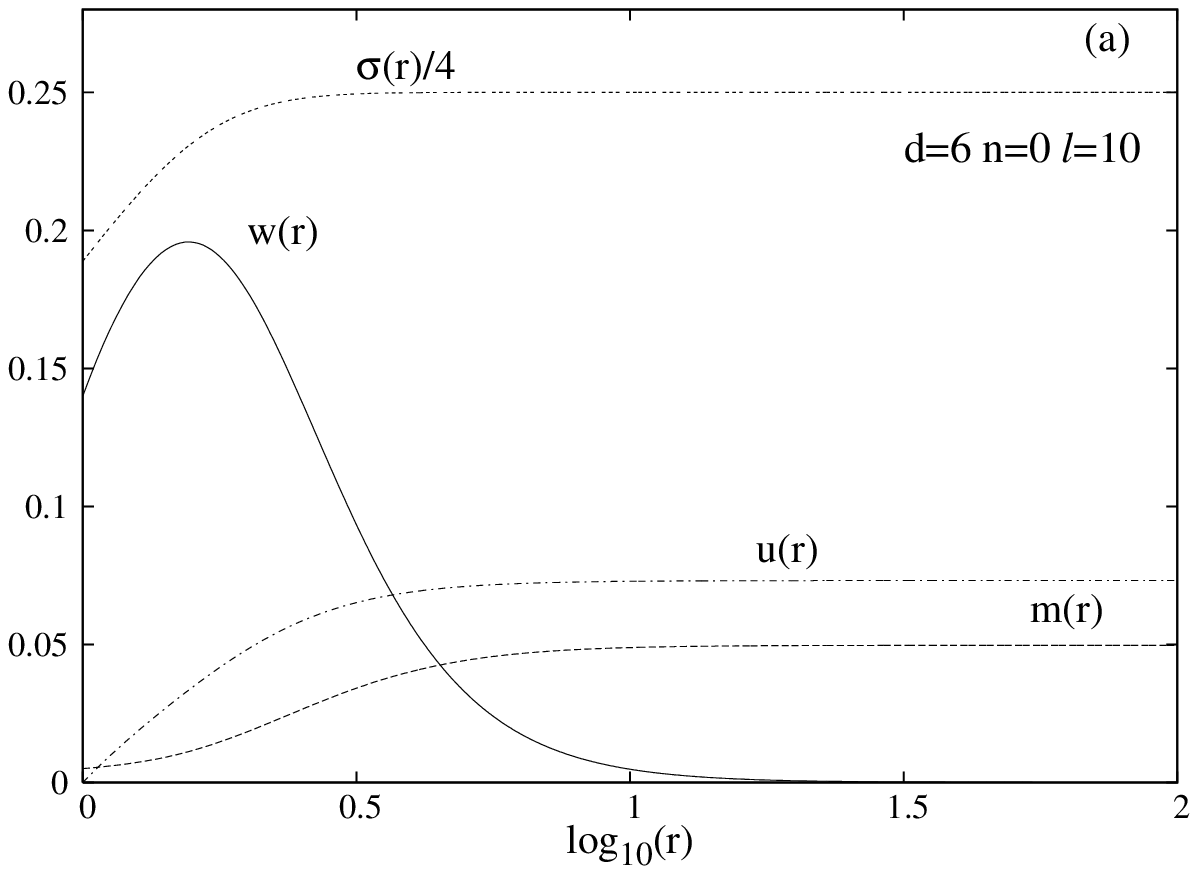}}
\hspace{5mm}%
 \resizebox{7cm}{6cm}{\includegraphics{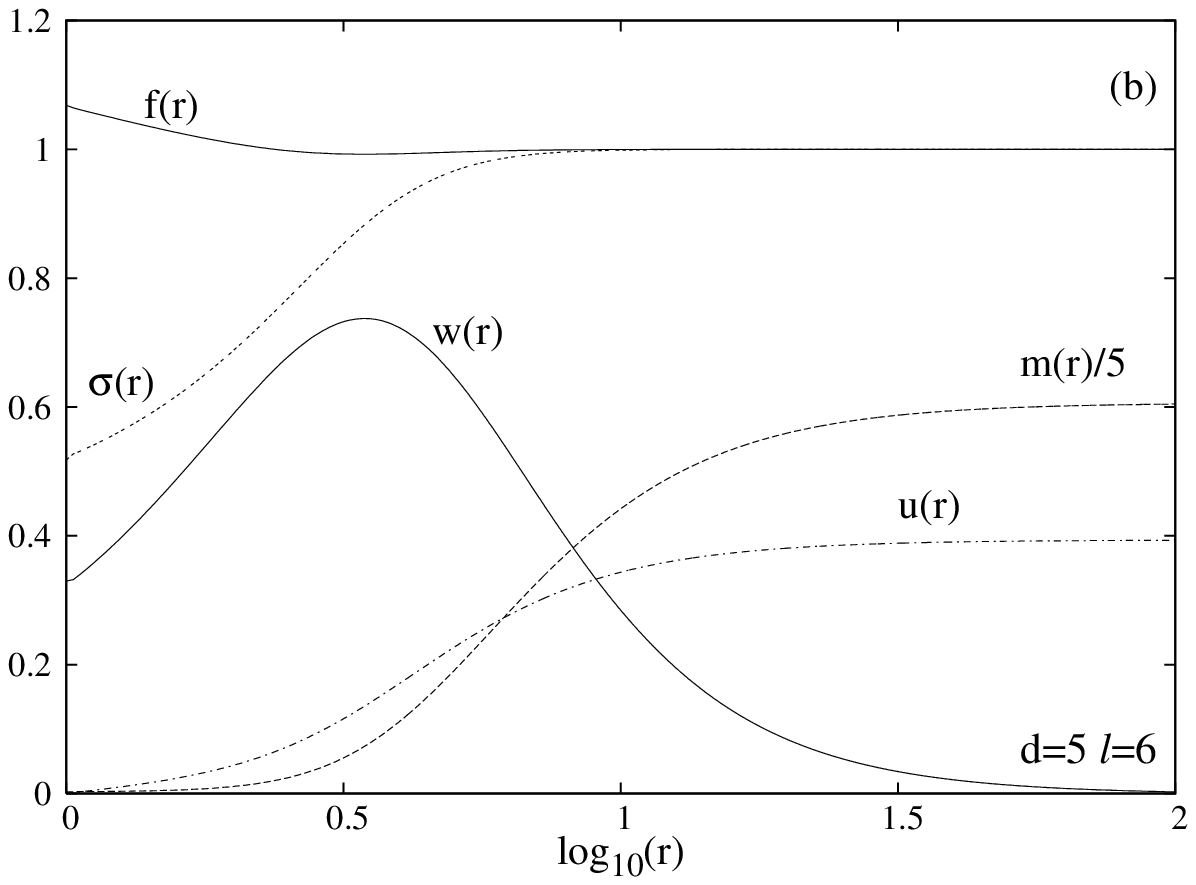}}	
%        \resizebox{7cm}{6cm}{\includegraphics{var-rh-d5k-1.eps}}	
\hss}
	\caption{ The  profiles of typical solutions are shown as a function of radial coordinate for Ansatz I (a)
and Ansatz II (b). }
\label{Fig25}
\end{figure}
%%%%%%%%%%%%%%%%%%%%%%%%%%%%%%%%%%%%
The constant $u_0$ in the asymptotic expansion (\ref{b2}) corresponds to the electrostatic potential  $\Phi=u_0/g$,
while $q$ fixes the electric charge density $Q_e= {(d-3)q}/{g}$.
Other quantities of interest are
the mass-energy density $M$, Hawking temperature $T$ and entropy density $S$,
\begin{eqnarray}
M=\frac{(d-2)M_0}{8\pi G},~~
T=\frac{N'(r_h)\sigma_h}{4\pi}=\frac{\sigma_h}{4 \pi}\left(\frac{(d-1)r_h}{\ell^2}-\frac{2m_1}{r_h^{d-3}}\right),~~
S=\frac{1}{4G}r_h^{d-2}.
\end{eqnarray}
The constant $J$ which enters the asymptotics of the magnetic
non Abelian potential $w(r)$
corresponds to an order parameter describing the deviation from the Abelian solution.

%%%%%%%%%%%%%%%%%%%%%%%%%%%%%%%%%%%%%%%%%%%%%%%%%%%%%%%%%%%%%%%
\section{Numerical solutions}
%%%%%%%%%%%%%%%%%%%%%%%%%%%%%%%%%%%%%%%%%%%%%%%%%%%%%%%%%%%%%%%
%%%%%%%%%%%%%%%%%%%%%%%%%%%%%%%%%%%%%%%%%%%%%%%%%%%%%%%%%%%%%%%
 \subsection{Scaling properties and general features}
%%%%%%%%%%%%%%%%%%%%%%%%%%%%%%%%%%%%%%%%%%%%%%%%%%%%%%%%%%%%%%%
We start by noticing that
 the equations (\ref{eqs}) are not affected by the transformation:
\begin{eqnarray}
\label{ss1}
 r\to \lambda r, ~~m\to \lambda^{d-3}m,~~\ell\to\lambda \ell,~~u\to \lambda u,~~\alpha\to  \alpha/\lambda
\end{eqnarray}
while $w,\sigma$ and $f$ remain unchanged. Thus, in this way  one can always take an arbitrary positive value for
$\alpha$. The usual choice is $\alpha=1$, which fixes\footnote{These are the units usually used in the literature on
EYM solutions~\cite{Volkov:1998cc}.
Note also that (\ref{ss1}) is a generic property
of the EYM system, shared by solutions with a different event horizon topologies.} the EYM length scale
$L=\sqrt{8\pi G/(g^2(d-2)}$, while the mass scale is fixed by ${\cal M}=(8\pi G/(g^2(d-2))^{(d-3)/2}/G$.
All other quantities get multiplied with suitable factors of $L$.
However, in what follows, to avoid cluttering our expressions with a complicated dependence of $(G,g,d)$,
we take a {\it unit} value for $\alpha$
and ignore the extra-factors of $g$ and $G$ in the expressions of various global quantities.

For solutions with a spherical event horizon, the event horizon radius and the value of the
magnetic potential on the horizon are independent parameters (see e.g. \cite{Bjoraker:2000qd}).
This is not the case for the solutions here, in which case
one can always set $r_h=1$ without any loss of generality.
This is a consequence of the following scaling symmetry of the system (\ref{eqs}):
\begin{eqnarray}
\label{ss2}
 r\to \lambda r,  ~~w\to \lambda w,~~u\to \lambda u, ~~
 m\to \lambda^{d-1}m,
\end{eqnarray}
while $\sigma,f$ and the cosmological constant remain unchanged\footnote{
The global quantities scale as follows:
$M\to \lambda^{d-1}M$, $T\to \lambda T$, $S\to \lambda^{d-2}S$,
$Q_e\to \lambda^{d-2}Q_e$, $\Phi \to \lambda \Phi$,
$J\to \lambda^{d-2}J$. }.

The system (\ref{eqs}) presents in addition two more scaling symmetries associated with
the functions $\sigma$ and $f$ ($e.g.$ $\sigma \to  \lambda \sigma $, $u\to  \lambda u$, $t\to  \lambda t$ etc.).
In the numerical procedure these symmetries are used to set  $\sigma(\infty)=f(\infty)=1$ and thus to fix
  the horizon values of the functions $\sigma$ and $f$.
Together with the other symmetries mentioned above, this leaves us with three numerically relevant parameters:
$w_h,u_1$ and the AdS length scale $\ell$. Since equations (\ref{eqs}) are invariant under the
transformation $w \rightarrow - w $, only values of $w _{h}>0$ are considered.

%
 %%%%%%%%%%%%%%%%%%%%%%%%%%%%%%%%%%%%
\begin{figure}[ht]
\hbox to\linewidth{\hss%
	\resizebox{7cm}{6cm}{\includegraphics{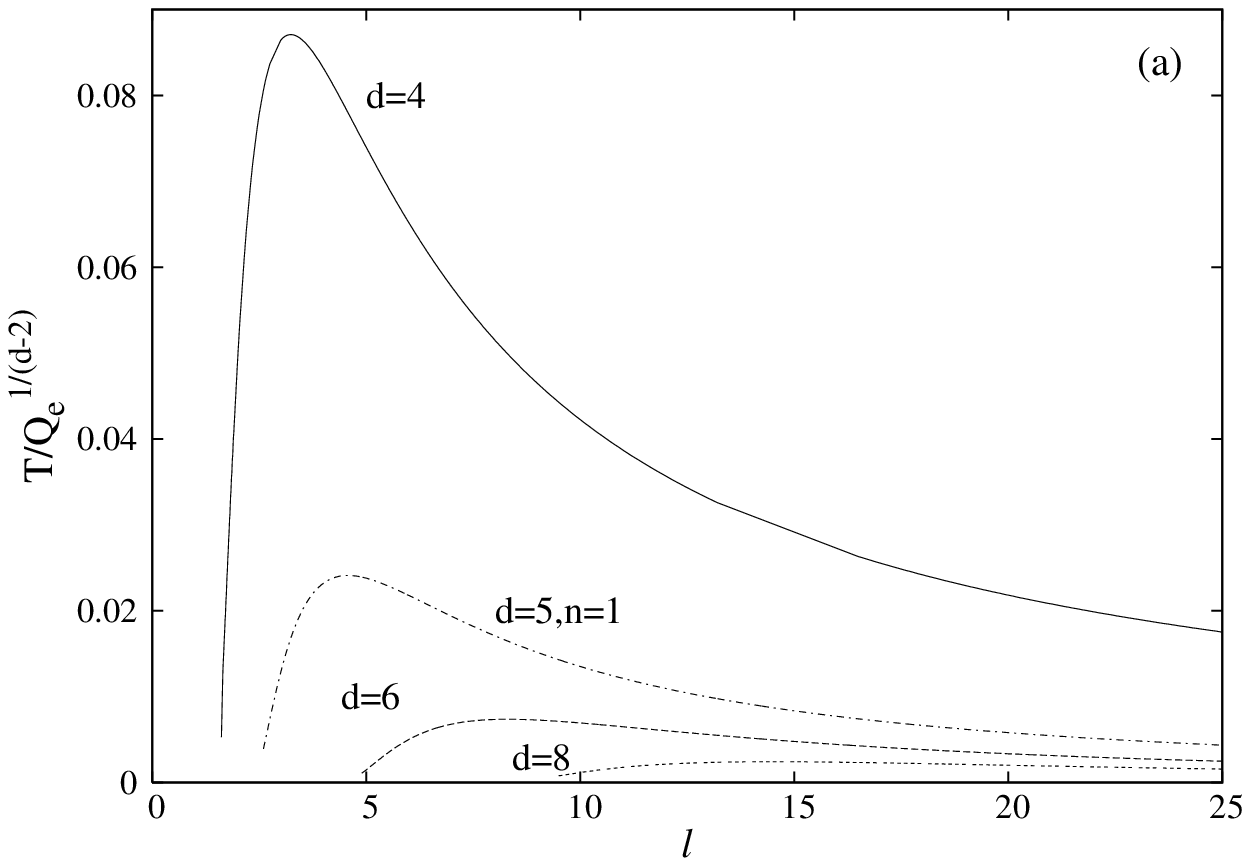}}
\hspace{5mm}%
 \resizebox{7cm}{6cm}{\includegraphics{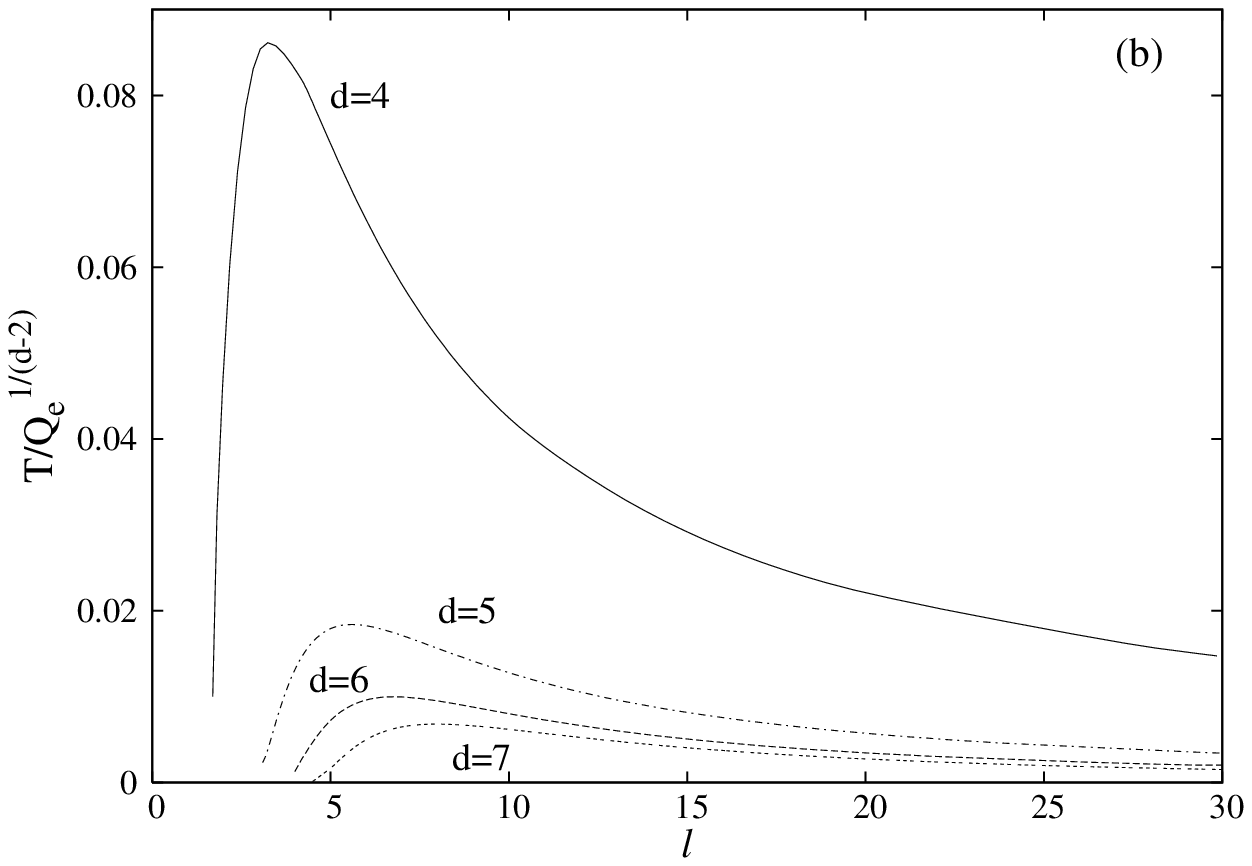}}	
%        \resizebox{7cm}{6cm}{\includegraphics{var-rh-d5k-1.eps}}	
\hss}
\caption{The phase diagram of the non Abelian solutions is plotted for several dimensions, for Ansatz I (a)
and Ansatz II (b). The value of $n$ is zero, except for the $d=5$ curve in Figure 2a.  }
\label{Fig25}
\end{figure}
%%%%%%%%%%%%%%%%%%%%%%%%%%%%%%%%%%%%

The equations  (\ref{eqs}) with boundary conditions implied   in turns  by (\ref{b1}), (\ref{b2}) have been
solved numerically, using a standard shooting method.
As expected, the properties of the solutions obtained for the
two distinct YM Ans\"atze (\ref{AnsatzI}), (\ref{AnsatzII}) are rather similar and thus we have
preferred to present them together. For the first case,
families of solutions have been constructed in a systematic way
for $d=4,6,8$  with $n=0$, and $d=5,7$ with $n=1$.
Several configurations with $d=6$, $n=2$ have been constructed as well.
When choosing instead the YM Ansatz (\ref{AnsatzII}), we have constructed solutions in $d=4,5$ and $6$ dimensions.
For every considered value of $\ell$, we could find regular black hole solutions
for only one interval $0\leq  w_h< w_h^c$.
The value of $w_h^c$ increases as $\ell$ decreases, $w_h=0$ corresponding to the RNAdS solution (\ref{RNAdS}) .

In all these cases, we noticed a number of common features.
The behaviour of solutions for generic initial data is such that $w\to w_0\neq 0$ at large $r$
(in which case the total mass-energy diverges), or else there is a singularity at finite $r$.
Given ($w_h,\ell$), solutions with the right asymptotic behaviour
(\ref{b2}) exist only for a discrete set of values of $u_1$.
As in the well known case of the Bartnik-McKinnon solutions~\cite{Bartnik:1988am}, the solutions here are also indexed
by the node number of the magnetic potential $w(r)$.  It turns out that
%However,
the configurations with nodes represent excited states whose energy is always greater
than the energy of the corresponding nodeless configurations, and are
therefore ignored in what follows.

For all solutions the functions $m(r)$, $\sigma(r)$ and $u(r)$ always increase monotonically
with growing $r$. However, $f(r)$ and $w(r)$ feature a more complicated behaviour.
Tyical solutions are presented in Figure 1 for both Ans\"atze. For sufficiently small $\omega_h$, all field variables remain
close to their values for the Abelian configuration with the same $r_h$.
Significant differences occur for large enough values of $\omega_h$
and the effect of the non Abelian field on the geometry becomes more and more pronounced.

 %%%%%%%%%%%%%%%%%%%%%%%%%%%%%%%%%%%%
\begin{figure}[ht]
\hbox to\linewidth{\hss%
	\resizebox{7cm}{6cm}{\includegraphics{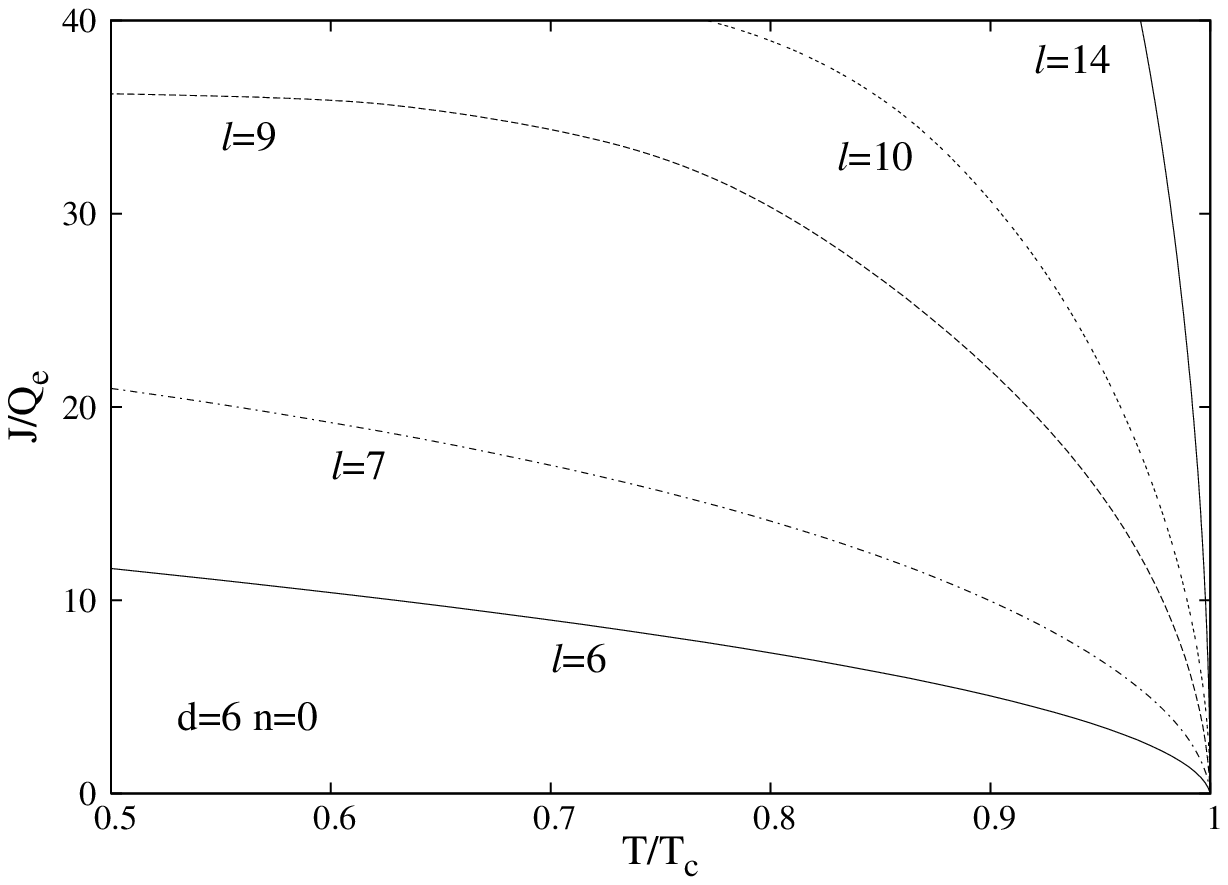}}
\hspace{5mm}%
 \resizebox{7cm}{6cm}{\includegraphics{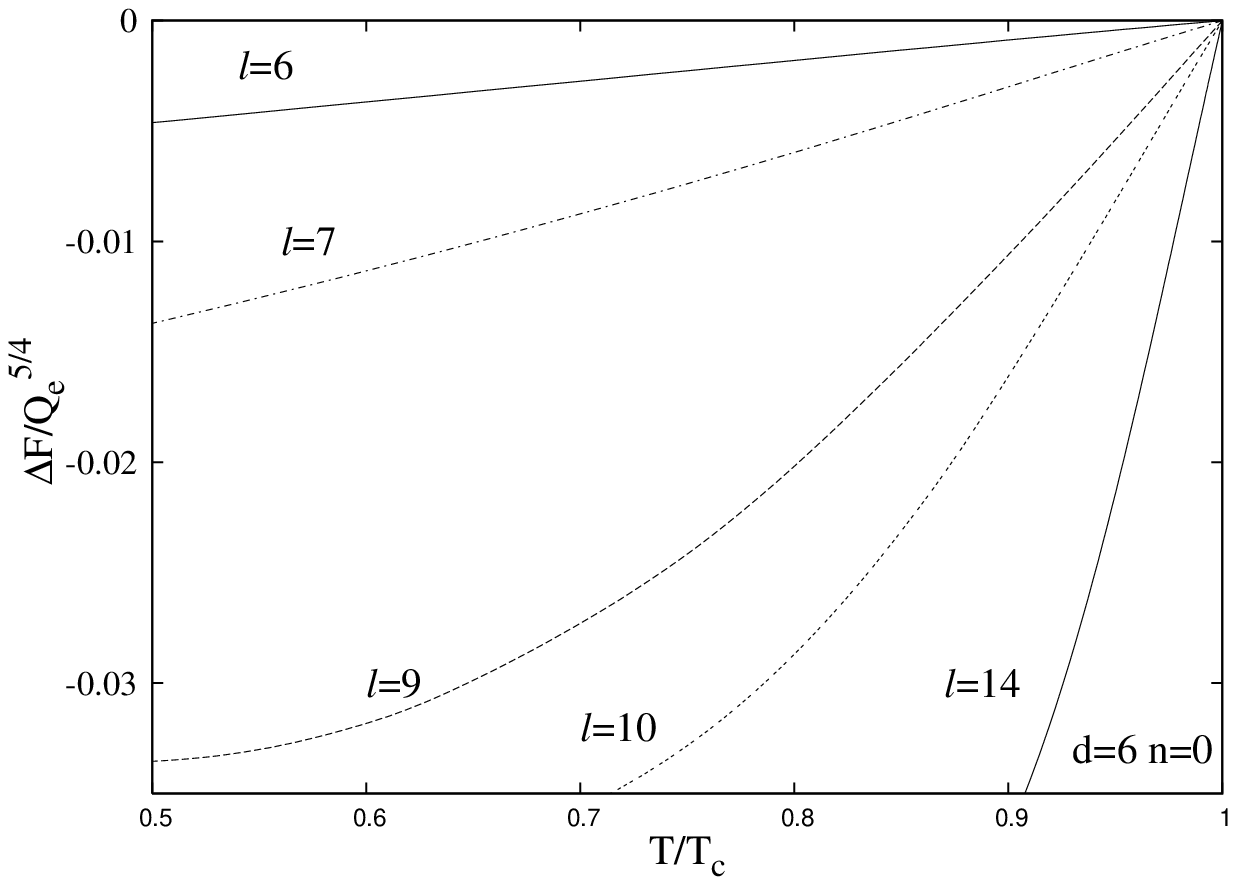}}	
%        \resizebox{7cm}{6cm}{\includegraphics{var-rh-d5k-1.eps}}	
\hss}
\hbox to\linewidth{\hss%
	\resizebox{7cm}{6cm}{\includegraphics{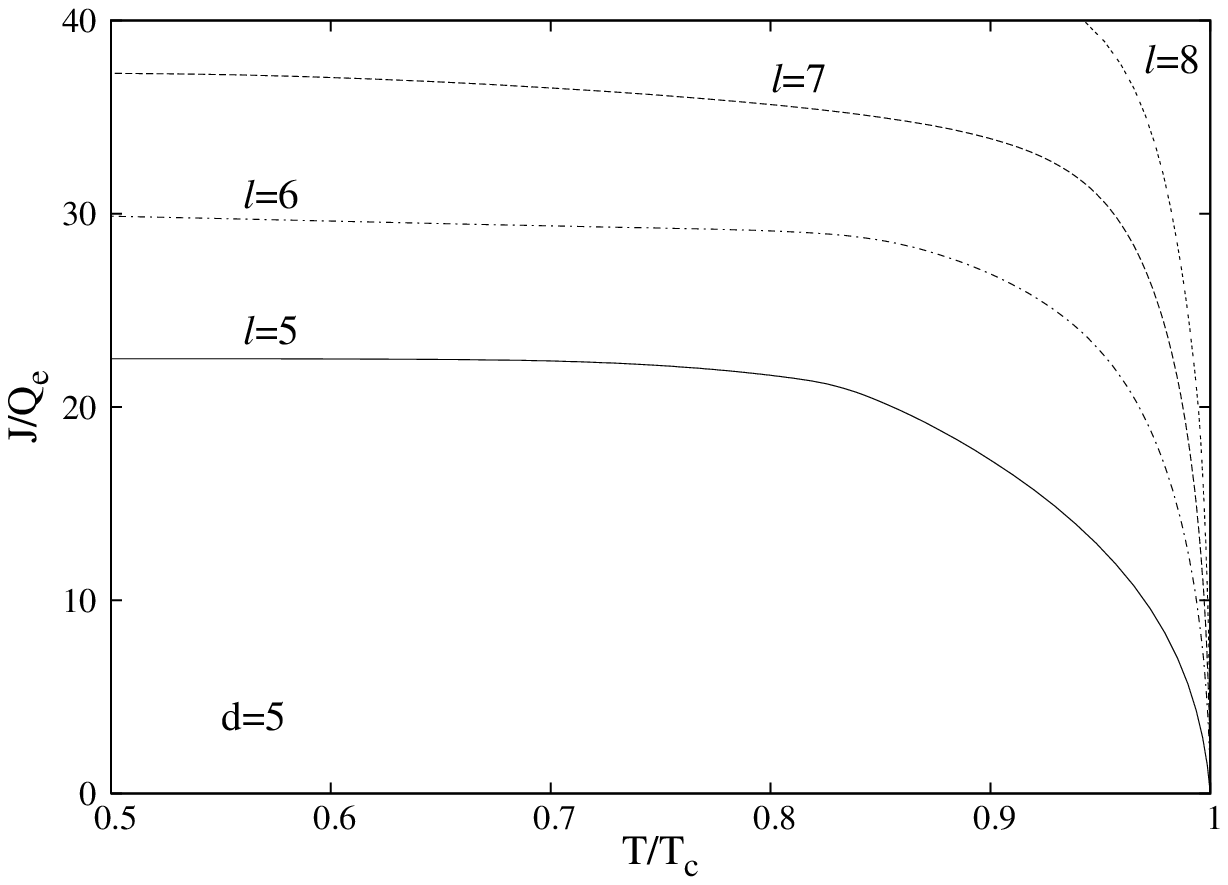}}
\hspace{5mm}%
 \resizebox{7cm}{6cm}{\includegraphics{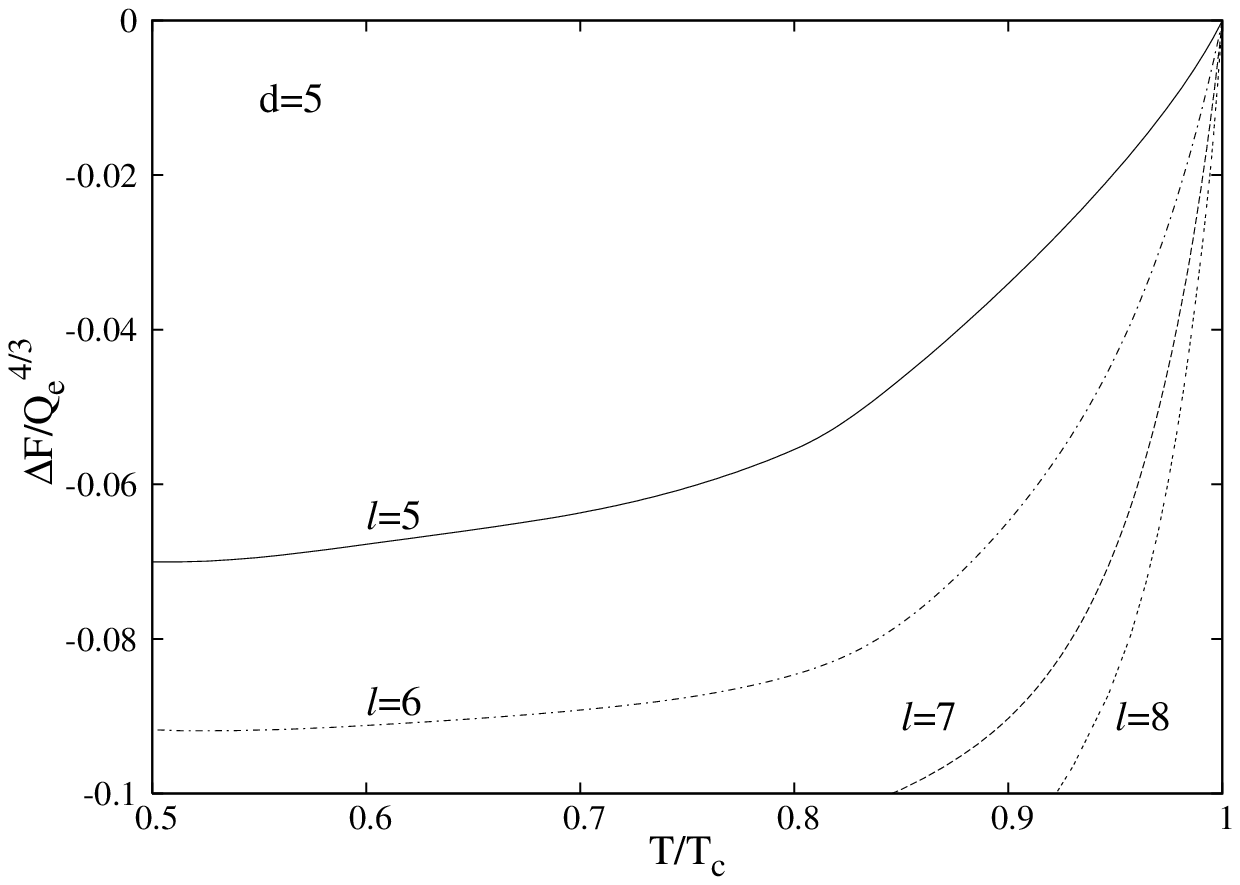}}	
%        \resizebox{7cm}{6cm}{\includegraphics{var-rh-d5k-1.eps}}	
\hss}
	\caption{ $J/Q_e$ and $\Delta F/Q_e^{(d-1)/(d-2)}$ are plotted as a function of $T/T_c$ for $d=5$ (Ansatz I) and
	$d=6$ (Ansatz II) non Abelian solutions.  }
\label{Fig25}
\end{figure}
%%%%%%%%%%%%%%%%%%%%%%%%%%%%%%%%%%%%

 %%%%%%%%%%%%%%%%%%%%%%%%%%%%%%%%%%%%%%%%%%%%%%%%%%%%%%%%%%%%%%%
 \subsection{Thermal properties and  superconducting horizons}
%%%%%%%%%%%%%%%%%%%%%%%%%%%%%%%%%%%%%%%%%%%%%%%%%%%%%%%%%%%%%%%
For all cases we considered, finite energy solutions
 were found only for values of the Hawking temperature less than a critical temperature
$T_c$. As in the $d=4$ case in \cite{Gubser:2008zu} this $T_c$ is, within numerical error, the temperature
at which the RNAdS solution admits a static linearised perturbation, with nonvanishing but infinitesimally
small\footnote{Our numerical code usually provided good quality solutions
for $T\gtrsim T_c/2$.} $w$. Also, our numerical results indicate that
$T_c$ goes to zero for some critical value of the AdS length scale $\ell$, but the corresponding
solutions do not appear to have a singular behaviour there.

These features are shown in Figure 2 for several values of $d$.
For a given dimension the part of the parameter space above the curve corresponds to
the unbroken phase, where only Abelian solutions exist.

In Figure 3, we plot several quantities which are invariant under the transformation (\ref{ss2}) as a function
of the ratio $T/T_c$. $\Delta F$ there is the difference in the free energy density, $M-TS$,
between a non Abelian solution and the the RNAdS solution with the same $T$ and $Q_e$.
As usual, $\Delta F < 0$ means that the non Abelian solution is thermodynamically favoured.

In all cases there is a second order phase transition with simple critical exponents, from RNAdS solutions to solutions
with normalisable non Abelian condensates. We have verified that for $T_m<T<T_c$, (with $T_m$  always around $0.7\ T_c$),
the solutions satisfy the universal relation
%\begin{eqnarray}
$ {J}/{Q_e}=j_{1/2}\sqrt{1-T/T_c}$,
%\end{eqnarray}
where $j_{1/2}$ depends on the model.

%%%%%%%%%%%%%%%%%%%%%%%%%%%%%%%%%%%%%%%%%%%%%%%%%%%%%%%%%%%%%%%
\section{Further remarks}
%%%%%%%%%%%%%%%%%%%%%%%%%%%%%%%%%%%%%%%%%%%%%%%%%%%%%%%%%%%%%%%
In this work we have presented arguments that the $d=4$ picture discovered in \cite{Gubser:2008zu}
is generic for the higher dimensional case as well.
Considering several values of $d\geq 5$, we have found evidence for the existence of a second order phase transition
with simple critical exponents, from the (electrically charged) RNAdS solutions with a flat event horizon
to non Abelian configurations with a nontrivial magnetic field.

 %
 %%%%%%%%%%%%%%%%%%%%%%%%%%%%%%%%%%%%
\begin{figure}[ht]
\hbox to\linewidth{\hss%
	\resizebox{7cm}{6cm}{\includegraphics{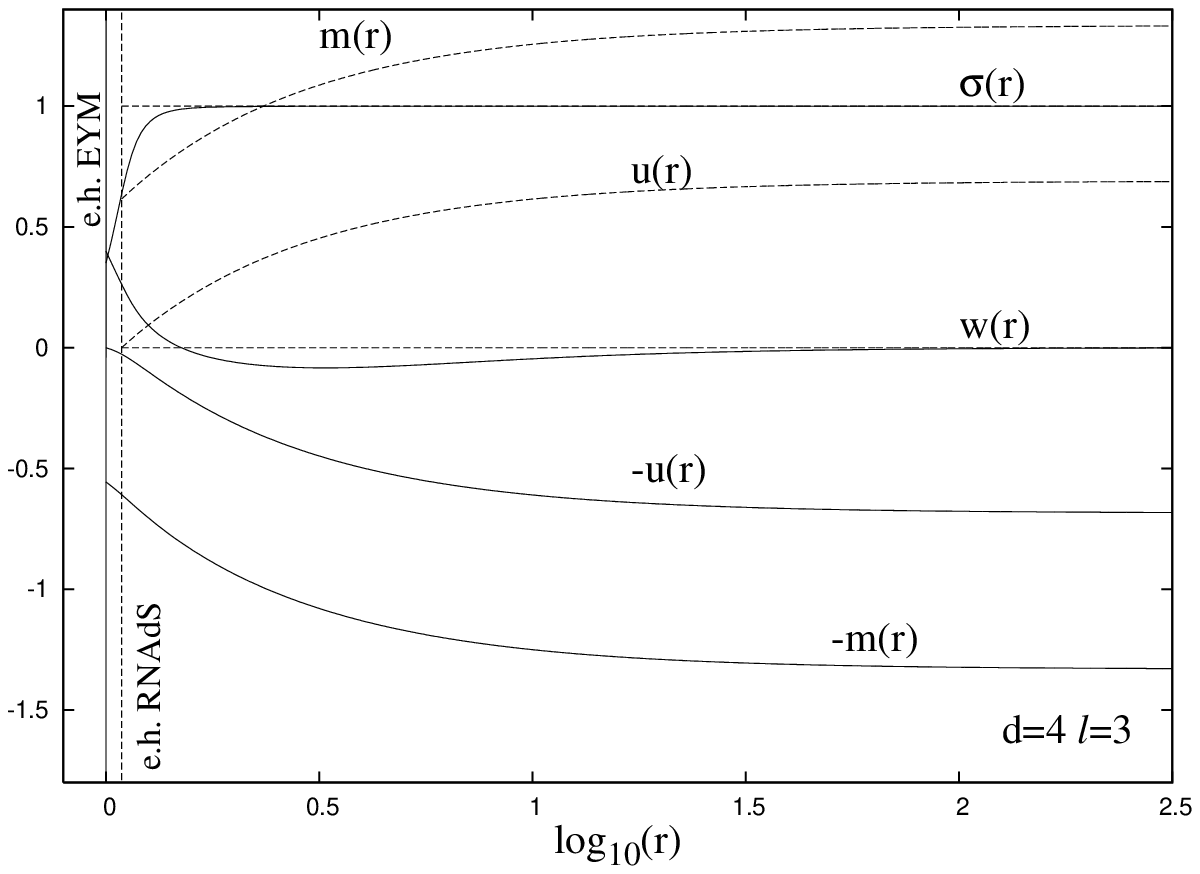}}
\hspace{5mm}%
 \resizebox{7cm}{6cm}{\includegraphics{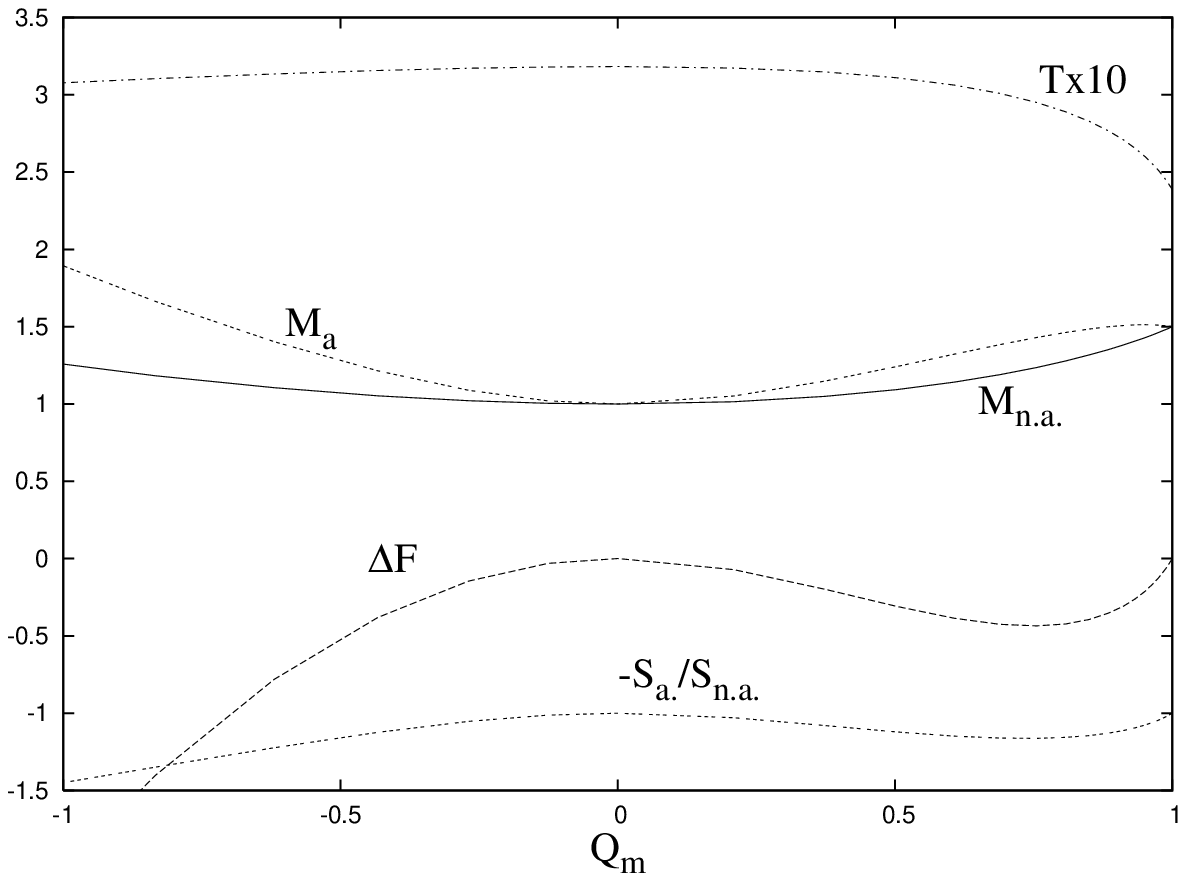}}	
%        \resizebox{7cm}{6cm}{\includegraphics{var-rh-d5k-1.eps}}	
\hss}
\caption{{\it Left}: The profiles of a $d=4$ dyonic non Abelian EYM solution with spherical horizon (continuous line)
is plotted together with a RNAdS
configuration with the same Hawking temperature and electric charge.
{\it  Right}: A number of quantities are presented as function of the magnetic charge for non Abelian monopoles and RNAdS
solutions with the same
temperature for $d=4$ solutions with $\ell=1$.
We have  ploted the Hawking temperature (multiplied with a factor of ten, for better visualisation), the ratio
between the entropies of the Abelian and the SU(2) solutions, the
black hole masses and the difference between the free energies.}
\label{Fig25}
\end{figure}
%%%%%%%%%%%%%%%%%%%%%%%%%%%%%%%%%%%%%%%

One should also note that the existence of these finite energy non Abelian solutions with $d>4$
represents a surprise in itself, since it contradicts the expectation based on the no-go theorems in
\cite{Okuyama:2002mh}, \cite{Volkov:2001tb}, \cite{Brihaye:2005tx},\cite{Radu:2005mj}.
The negative results in the latter were proven for configurations with a spherical topology of the event horizon,
in which case the electric potential necesarily vanishes when $d\ge 5$.
Purely magnetic EYM solutions with finite mass were found by considering
corrections to the YM Lagrangian consisting in higher order terms of the
Yang--Mills hierarchy of the form $L_p={\mbox Tr\ }F(2p)^2$
where $F(2p)$ is the $2p$-form $p$-fold totally antisymmetrised product
of the $SO(d)$ YM curvature $2$-form $F(2)$
(see e.g. \cite{Brihaye:2002hr} for asymptotically flat configurations and \cite{Radu:2005mj}, \cite{Brihaye:2006xc} for solutions
with a cosmological constant). Such systems occur in the
low energy effective action of string theory and are, in some sense, the non Abelian counterparts of the
Lovelock gravitational
hierarchy~\footnote{However, inclusion of members of this gravitational hierarchy
turns out to be of no practical utility because subject to the symmetries imposed such terms with the requisite
scaling properties vanish.}.
By contrast, we have verified that to construct finite energy EYM solutions with Ricci flat
horizons, inclusion of higher derivative terms in the YM curvature to the action (\ref{action})
is not necessary. For this reason we have restricted our attention to the usual gravitating YM system,
as inclusion of higher order YM terms results only in (unimportant) quantitative changes.

An important question is, whether it is only for solutions planar event horizon that the non Abelian solution is
themodynamically favoured over the Abelian one? To answer this question, consider simply the case of the $d=4$ dyonic
SU(2) black holes with spherical event horizon topology originally discussed in Ref. \cite{Bjoraker:2000qd}.
These solutions are found within the Ansatz
\begin{eqnarray}
\label{hoso}
&&ds^2=\frac{dr^2}{N(r)}+r^2(d\theta^2+\sin^2\theta d\phi^2)-\sigma^2(r)N(r)dt^2,
~~~{\rm with~~}~N(r)=1-\frac{2m(r)}{r}+\frac{r^2}{\ell^2},
\\
\nonumber
&&A=\frac{1}{2} \Big\{u(r)\tau_3dt+ w(r)\tau_1d\theta
+(\cot\theta\tau_3+w(r)\tau_2)\sin\theta d\varphi\Big\},
\end{eqnarray}
where  $\tau_a$ are the Pauli matrices.
Without any loss of generality, by using the symmetry (\ref{ss1}), one can set $4 \pi G=g=1$.
The problem reduces in this case to a system of four coupled ordinary differential equations.
The properties of these solutions including the boundary conditions
and the asymptotic expansion can be found\footnote{The function $p(r)$ in \cite{Bjoraker:2000qd} corresponds to
$1/\sigma(r)$ in the Ansatz (\ref{hoso}).} $e.g.$ in \cite{Bjoraker:2000qd}.
The generic behaviour of the solutions is such that they have a nonvanishing magnetic charge $Q_m=1-w^2(\infty)$.
Non Abelian solutions with $Q_m=1$ are found for special values of $u'(r_h)$ and have $g_{rr} g_{tt}\neq -1$.
The dyonic Abelian RNAdS solution with unit magnetic charge corresponds to
$ w(r)=0,~u(r)=u_0+q/r,~\sigma(r)=1,~ m(r)=M_0-(1+q^2)/2r$.
Considering again the case of a canonical ensemble, we have found numerical evidence for the existence of non Abelian
solutions which are thermodynamically favoured over the Abelian ones.
An example of such a situation is presented in Figure 4 for an AdS length scale $\ell=3$.
There, the Hawking temperature and the electric charge are $T\simeq 0.0053$ and
$Q_e\simeq 0.751$ for both solutions,
while the masses are slightly different: $M$(RNAdS)$\simeq 1.334$ and $M\simeq 1.33$ for the non Abelian counterpart.
This implies $\Delta F<0$ and thus the existence of a phase transition\footnote{
Note also that, for these parameters there exists only one Abelian configuration (this is a property of all symmetry
breaking solutions we have found so far). However,
as discussed in \cite{Chamblin:1999hg}, the general picture is much more complicated,
with the possible existence of several branches of Abelian configurations.
Moreover, similar to the case of black holes with a flat horizon, thermodynamically favoured non Abelian
solutions seem to exist only for a limited region of the parameter space.}.

One should also remark that, for $d=4$, the presence of a non Abelian electric field is
not crucial for the existence of a phase transition between Abelian and non Abelian solutions.
Setting $u(r)=0$ in the Ansatz (\ref{hoso}),
our  numerical results indicate the existence of purely magnetic non Abelian configuration which are
thermodynamically favoured
over the abelian solutions with the same magnetic charge\footnote{Due to the existence of electric-magnetic
duality in $d=4$ Einstein-Maxwell theory, one can consider electrically charged RNAdS solutions as well.}
and temperature. In Figure 4b we plot a number of relevant quatities for a family of $d=4$ EYM monopole black holes
with $r_h=1$ and $\ell=1$ and the corresponding RNAdS solutions.
The gauge potential $w(r)$ is nodeless for all solutions there. The solution with $Q_m=0$ corresponds to the
Schwarzschild-AdS (SAdS) black hole with a spherical horizon. One can see that for $\Lambda=-3$
all non Abelian solutions with $r_h=1$ have $\Delta F<0$.
However, the generic picture is more complicated, with a nontrivial dependence on $\ell,r_h$.

A study of these aspects is beyond the purposes of this work and will be presented elsewhere.

We close by remarking that the asymptotic AdS structure of the spacetime is crucial for
the existence of such solutions.
As proven in \cite{bizon} for $d=4$, the asymptotically flat EYM solutions have no magnetic charge while their
electric part vanishes identically.
Moreover, by using the data in  \cite{Volkov:sv},
one can easily verify the difference between the free energy of a SU(2) hairy black hole
and the Schwarzschild solution with the same temperature is always positive.
 \\
\\
{\bf Acknowledgement}
\\
This work is carried out in the framework of Science Foundation Ireland (SFI) project
RFP07-330PHY.
 The work of ER was supported by a fellowship from the Alexander von Humboldt Foundation.

%%%%%%%%%%%%%%%%%%%%%%%%%%%%%%%%%%%%%%%%%%%%%%%%%%%%%%%%%%%%%%%%%%%%%%%%%%%%%%

%%%%%%%%%%%%%%%%%%%%%%%%%%%%%%%%%%%%%%%%%%%%%%%%%%%%%%%%%%%%%%%%%%%%%%%%%%%%%%

%%%%%%%%%%%%%%%%%%%%%%%%%%%%%%%%%%%%%%%%%%%%%%%%%%%%%%%%%%%%%%%%%%%%%%%%%%%%%%

\end{document}